\documentstyle[multicol,aps,epsf]{revtex}
\begin{document}
\input psfig.tex
\draft
\title{A Model for Growth of Binary Alloys with Fast Surface Equilibration}
\author{Barbara Drossel and Mehran Kardar}
\address{Physics Department, Massachusetts Institute of Technology, Cambridge,
MA 02139, USA}
\date{\today}
\maketitle
\begin{abstract}
We study a simple growth model for $(d+1)$-dimensional films of binary alloys
in which atoms are allowed to interact and equilibrate at the surface, but
are frozen  in the bulk. The resulting crystal is highly {\it anisotropic}:
Correlations perpendicular to the growth direction are identical to
a $d$-dimensional {\it two-layer system in equilibrium},
while parallel correlations generally reflect the (Glauber) dynamics of such a
system.
For stronger in--plane interactions, the correlation volumes change from oblate
to highly prolate shapes near a critical demixing or ordering transition.
In $d=1$, the critical exponent $z$ relating the scaling of the two correlation
lengths varies continuously with the chemical interactions.
\end{abstract}

\begin{multicols}{2}
\section{Introduction}

Growth by vapor deposition is a highly effective process for producing high
quality materials. The resulting films can show properties that are very
different from systems in bulk equilibrium \cite{fro91,roo95}.
For example, in some binary alloys\cite{roo95}, the deposited atoms
are highly mobile as long as on the surface, but relatively immobile once
incorporated in the bulk. As a result, the surface fluctuations that are
formed during the growth process are frozen into the bulk. A characteristic
signature of such (metastable) phases is {\it anisotropic} correlations that
are
related to the growth direction, and are absent in bulk equilibrium.

A number of models for composite film growth have been introduced in
the past\cite{wel73,ent77,kim85,dav87,YBY90,roo93}.
Generally in these models, the probability that an incoming atom sticks to a
given surface site depends on the state of neighboring sites in the layer
below.
Once a site is occupied, its state does not change any more, and thus the
surface configuration becomes frozen in the bulk.
Such growth rules are equivalent to (stochastic) cellular automata, where
each site is updated in parallel as a function of the states of its neighbors.
Subsequent states of the cellular automaton correspond to successive
layers in the crystal.

It is in general not possible to calculate exact correlation functions for
such (non-equilibrium) growth processes.
The exception occurs in special cases where the growth rules satisfy
a detailed balance condition, relating their stationary behavior to an
equilibrium system of one lower dimension\cite{gri85}.
However, it can be shown that if $d$-dimensional probabilistic
cellular automata with two states, and up-down symmetry, undergo a
symmetry breaking, their critical behavior is identical to the corresponding
Ising model in equilibrium \cite{gri85}. Correlations in time are then
equivalent to
those generated by Glauber dynamics of the Ising system.
$(d+1)$-dimensional crystals grown according to the rules of these cellular
automata therefore have an order-disorder phase transition with correlations
perpendicular to the growth direction characterized by the critical exponent
$\nu$, and those parallel to the growth direction by the exponent $\nu z$ of
the $d$-dimensional Ising model ($z$ being the appropriate dynamical
critical exponent).

Since in the above models the local adsorption probabilities depend on the
surface states, some form of fast redistribution of atoms is implied.
One possibility is the mixing of particles in the gas phase prior to
adsorption.
However, it is easy to envision conditions where the local adsorption
probabilities are determined by densities in the gas phase (e.g. in
ballistic deposition). In such cases, rapid desorption from unfavorable
locations on the surface may provide the appropriate redistribution
mechanism.  However, given the high mobility of particles on the surface,
surface diffusion is another important process. In this case, it is essential
to also include the interactions between the diffusing surface particles
which eventually leads to formation of domains and islands.

In this paper, we include the interactions between atoms on the top layer,
which is assumed to equilibrate completely (by surface diffusion {\it or}
desorption--resorption mechanisms) before another layer is added.
Such an assumption is realistic only if the growth rate is much slower
than characteristic relaxation times of the surface layer. Its limits of
validity are discussed in the conclusion; in particular, it is likely
to break down in the vicinity of a critical point.
We show that this model satisfies detailed balance, and can therefore
be analyzed with methods from equilibrium statistical physics.
While the resulting critical behavior is similar to the previously studied
(cellular automata) models, the `temporal' correlations in 1+1 dimension,
are very different from those obtained from the mapping to Glauber models:
On approaching zero temperature, the diverging vertical and horizontal
correlations lengths are related by exponent $z$ that is in general larger
than the Glauber value of 2, and varies continuously with the ratio of the
perpendicular and parallel coupling constants.
In all dimensions, if the ratio between the two coupling constants is larger
than unity, the two correlation lengths cross at some temperature,
changing the shapes of typical correlated clusters from oblate (at
high growth temperatures) to prolate (near the critical point).

\section{Model}
\label{model}

The model is defined as follows: We consider a $d+1$ dimensional hypercubic
lattice and two kinds of atoms, $A$ and $B$. Let $\epsilon_{AA}$,
$\epsilon_{AB}$,
and $\epsilon_{BB}$ be the interaction energies between neighboring atoms of
types $AA$, $AB$, and $BB$ respectively. When
each layer has $N$ sites, there are $2^N$ possible configurations for a layer.
The energy cost for adding a layer of configuration $\gamma$ on top of one with
configuration $\alpha$ is the sum of the internal energy $E_{\gamma}$ of the
new layer, and of the interaction energy $V_{\alpha \gamma}$ with the previous
layer. These energies are just the sums of all local bonds $\epsilon_{ij}$
between
nearest neighbors $ij$ within the new layer and between
the two layers respectively. In addition, $E_\gamma$ contains  a chemical
potential $\mu_A N_A + \mu_BN_B$ that is related to the partial pressures of
$A$ and $B$ atoms in the gas phase.

Since we assume that the top layer is in thermodynamic equilibrium, the
conditional probability that it is in configuration $\gamma$, given
configuration
$\alpha$ for the layer below, is
\begin{equation}
W_{\gamma \alpha} = {\exp\left[-\beta (E_{\gamma} + V_{\alpha\gamma}) \right]
\over \sum_{\delta} \exp\left[-\beta (E_{\delta} + V_{\alpha\delta})
\right]}\, , \label{W}
\end{equation}
where $\beta = 1/k_BT$; $T$ is the temperature at which the crystal is grown,
and $k_B$ is the Boltzmann constant.
After adding many layers, the probability for finding a given configuration
$\gamma$ is determined by the stationarity condition
\begin{equation}
P_{\gamma} = \sum_{\alpha} W_{\gamma\alpha} P_\alpha\,,
\label{stationary}
\end{equation}
which has the solution
\begin{eqnarray}
P_\alpha &= &{\sum_\gamma \exp\left[-\beta(E_\alpha + E_\gamma +
V_{\alpha\gamma})\right] \over \sum_{\delta, \nu} \exp\left[-\beta (E_\delta
+ E_\nu + V_{\delta\nu})\right] }\nonumber \\
&\equiv& {\sum_\gamma \exp\left[-\beta H_{\alpha\gamma}\right] \over
\sum_{\delta, \nu} \exp\left[-\beta H_{\delta\nu}\right] }\, .
\label{P}
\end{eqnarray}
The above expression is the equilibrium probability
for the top layer of a two-layer system, obtained after summing over the
states of the bottom layer.
Transverse correlation functions (i.e. perpendicular to the growth direction)
are therefore exactly the same as correlation functions in a two-layer system.

In fact, from Eqs.~(\ref{W}) and (\ref{P}) it easily follows that the system
satisfies detailed balance, i.e.,
\begin{equation}
W_{\alpha\gamma}P_\gamma = W_{\gamma\alpha}P_\alpha.
\label{detailedbalance}
\end{equation}
This means that (beyond a transient thickness) the crystal looks the same along
or against the growth direction, and that the sequence of layers corresponds to
time evolution of thermodynamic equilibrium states.
This generalizes the previous results for cellular automata, which are obtained
by setting the in--plane interactions $E_\alpha$ to zero.
As in such cellular automata, the $(d+1)$-dimensional system has
transverse properties like $d$-dimensional models. In particular, we expect
phase transitions  to occur at the same temperature as
for a $d$-dimensional two-layer system.

The two-layer Hamiltonian occurring in Eq.~(\ref{P}) can be rewritten in terms
of Ising variables by introducing the spin states $\sigma = +1$ for an $A$
atom, and $\sigma = -1$ for a $B$ atom. Using
$ J = (\epsilon_{AA} - 2\epsilon_{AB} + \epsilon_{BB})/4$ and $h =
3(\epsilon_{AA} - \epsilon_{BB})/4 + (\mu_A-\mu_B)/2$, and neglecting an
additive constant, the two-layer Hamiltonian becomes
\begin{eqnarray}
H_{\alpha\gamma} &=& J_\perp\sum_{<i,j>}
\left(\sigma_i^{(\alpha)}\sigma_{j}^{(\alpha)} +
\sigma_i^{(\gamma)}\sigma_{j}^{(\gamma)} \right) \nonumber \\
&+&  \sum_{i=1}^N \left(J_\parallel\sigma_i^{(\alpha)}\sigma_{i}^{(\gamma)}+
h\sigma_i^{(\alpha)} + h\sigma_i^{(\gamma)}\right)\,,
\label{H}
\end{eqnarray}
where we have also allowed for the possibility of anisotropic couplings
parallel and perpendicular to the growth direction.
As a function of  the ``field'' $h$, there is a first-order transition between
A and B rich phases. Here, we focus on the coexistence line at $h=0$, which
terminates at a critical point, where the correlation lengths diverge
both parallel and perpendicular to the growth direction.
For $h=0$, the Hamiltonian is invariant under the reversal of the signs of all
spins. This corresponds to the situation where an equal number of $A$ and $B$
atoms are deposited. If the coupling constants $J_\perp$ and $J_\parallel$ are
positive, atoms of the
same kind attract each other, and the two types of atoms phase separate below
the critical temperature. If $J_\perp$ and $J_\parallel$ are both negative, and
if the lattice structure allows a unique ground state, the low-temperature
phase is
``antiferromagnetic'' with $A$ and $B$ atoms sitting on different sublattices.
These two cases can be mapped onto each other by reversing the sign of the
coupling constants, and
of the spins on one of the sublattices. If $J_\parallel < 0$ and $J_\perp > 0$,
the low-temperature phase alternates between $A$ and $B$ layers. In the
opposite case, all layers have an antiferromagnetic occupation in the ground
state, with sublattices for the same type of atoms laying on top of each other.
Since these situations can also be mapped to the ferromagnet, we shall focus
on $J_\parallel,J_\perp >0$ from now on.

\begin{figure}
\narrowtext
\centerline{\epsfysize=2.5in 
\epsffile{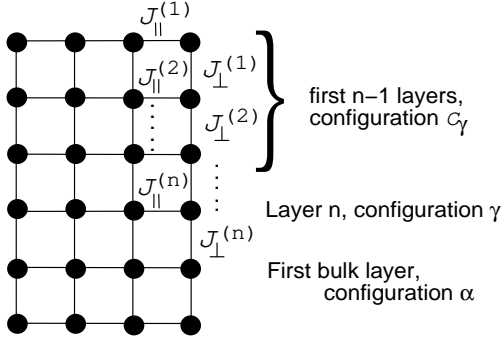}}
\caption{A system where the top $n$ layers equilibrate.}
\label{nlayers1}
\end{figure}

\begin{figure}
\narrowtext
\centerline{\epsfysize=2.5in 
\epsffile{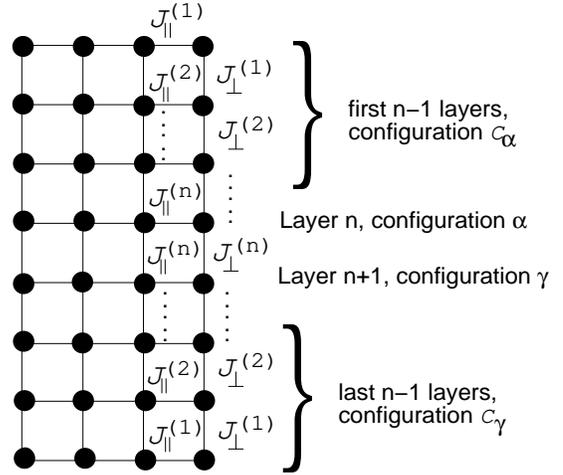}}
\vskip 0.5cm
\caption{Graphical illustration of the stationary distribution $P_\gamma$.}
\label{nlayers2}
\end{figure}

Generalizing our model, by allowing several equilibrating layers at the 
surface to relax, is straightforward. To mimic the large energy of the
impinging particles, as well as their modified environment, we can assign 
each of the top $n$ layers from the surface a different temperature, 
depending on its depth. Equivalently, we can give to each layer different 
interaction parameters (see Fig.~\ref{nlayers1}). 
The probability that a layer with configuration 
$\gamma$ follows one in configuration $\alpha$ in the bulk is obtained 
by considering the layer at the moment when it is the $n$th layer from 
the top, i.e. immediately before its configuration is frozen.
Denoting the configuration of the first $n-1$ layers by ${\cal C}_\gamma$ 
and their energy (including the coupling to the $n$th layer,
and different interaction constants in the different layers)
by $E({\cal C}_\gamma)$, the conditional probabilities $W_{\gamma\alpha}$
 can be written as
\begin{displaymath}
W_{\gamma \alpha} = {\sum_{{\cal C}_\gamma}\exp\left\{-\beta [E_{\gamma} +
V_{\alpha\gamma} + E({\cal C}_\gamma)] \right\}
\over \sum_{\delta,{\cal C}_\delta} \exp\left\{-\beta [E_{\delta} +
V_{\alpha\delta}+ E({\cal C}_\delta)]
\right\}}\, .
\end{displaymath}
Following the approach for the case $n=1$, we can show that the set of weights
\begin{displaymath}
P_\alpha = {\sum_{\gamma,{\cal C}_\gamma,{\cal C}_\alpha} \exp\left\{-\beta[E_\alpha + E_\gamma
+ V_{\alpha\gamma} + E({\cal C}_\gamma) + E({\cal C}_\alpha)]\right\}\over
\sum_{\delta, \nu,{\cal C}_\delta,{\cal C}_\nu } \exp\left\{-\beta [E_\delta
+ E_\nu + V_{\delta\nu}+ E({\cal C}_\delta)+ E({\cal C}_\nu)]\right\}}
\end{displaymath}
describe a stationary state. It is easy to verify that this stationary 
solution satisfies detailed balance. 
The  stationary state corresponds to an
equilibrium Hamiltonian with $2n$ layers, and
interactions which depend on the distance from the closest surface (see 
Fig.~\ref{nlayers2}). 
In this symmetric system, configurations
of  the layers $n$ and $n+1$ have the same probability distribution. Layer
$n$ of Fig.~\ref{nlayers2} therefore sees exactly the same environment as 
layer $n$ of  Fig.~\ref{nlayers1} in a grown bulk (beyond a transient
thickness). 
The top (and also the bottom)  layer of Fig.~\ref{nlayers2}
describe the equilibrium at the top surface layer, while the $n$th layer 
describes
the transverse correlations in the bulk. While the correlations
parallel to the growth direction will be more complicated, the
general conclusions derived below for the case 
$n=1$ should remain valid. In the following, we restrict our discussion
 to the case $n=1$. 

\section{One Dimension}

The $(1+1)$-dimensional model can in fact be solved exactly, and displays
interesting critical properties.
The two-spin correlations perpendicular to the growth direction are
obtained from the equivalent ladder of spins as
\begin{eqnarray}
g_{\perp}(l) &=& \langle
\sigma_i^{(\alpha)}\sigma_{i+l}^{(\alpha)}\rangle\nonumber\\
&=&\sum_{\alpha}\sigma_i^{(\alpha)}\sigma_{i+l}^{(\alpha)}{\sum_\gamma
\exp\left[-\beta H_{\alpha\gamma}\right] \over \sum_{\delta, \nu}
\exp\left[-\beta H_{\delta\nu}\right] }\nonumber\\
&\propto& \left( {\lambda_2 \over \lambda_1} \right)^l\, ,
\label{corr}
\end{eqnarray}
where $\lambda_1$ are $\lambda_2$ are the largest and next largest
eigenvalues of the $4\times 4$ transfer matrix. Explicit diagonalization of
the matrix gives
\begin{eqnarray}\label{evs}
\lambda_1 &=& \exp(\beta J_\parallel + 2\beta J_\perp) + \exp(\beta J_\parallel
- 2\beta J_\perp )\nonumber\\
&&+ {\cal O}(\exp(-\beta J_\parallel - 2\beta J_\perp))\,, \nonumber\\
\lambda_2 &=& \exp(\beta J_\parallel + 2\beta J_\perp) - \exp(\beta J_\parallel
- 2\beta J_\perp )\,.
\end{eqnarray}
It is then easy to check that in the limit of zero temperature
($\beta \to\infty$), the correlations decay exponentially with
a  correlation length $\xi_{\perp}$ which diverges as
\begin{equation}
 \xi_{\perp} \simeq \exp\left(4\beta J_\perp\right)/2\,.
\label{xiperp}
\end{equation}

\begin{figure}
\narrowtext
\centerline{\epsfysize=3in 
\epsffile{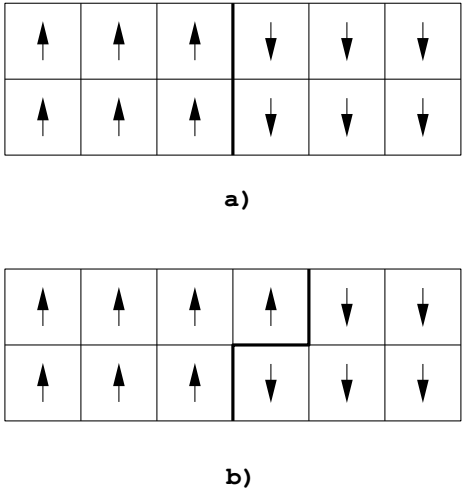}}
\caption{a) Straight domain walls are the lowest energy excitations of
a two layer system.
b) A  kink in the domain wall cost additional energy.}
\label{wall}
\end{figure}

The following argument provides a better physical understanding for the
form of $\xi_{\perp}$: At low temperatures, the energy of the system is only
slightly above its ground state, and consequently most of the spins are
parallel. The lowest-energy excitations of the ground state that destroy
long range correlations are straight domain
walls that cost an energy of $4J_\perp$, and occur with a probability
$2\exp(-4\beta J_\perp)$
(see Fig.~\ref{wall}). 
The factor two is included since either the
left part or the right part of the system can flip in order to create a domain
wall at a given location. More complicated excitations like domain walls with
steps cost at least an energy of $4J_\perp + 2 J_\parallel$ and occur with
probabilities smaller by a
factor of $\exp(-2\beta J_\parallel)$ than straight domain walls. The
transverse
correlation length is given by the inverse density of domain walls, leading to
Eq.~(\ref{xiperp}).

The correlation length parallel to the growth direction can be obtained by
similar considerations. Parallel correlations are destroyed by the motion of
domain walls. The probability that a domain wall moves by one lattice site to
the right or to the left when a new layer is added is $\exp(-2\beta
J_\parallel)$ (see
Fig.~\ref{wall}). Since these steps in the domain wall are independent from
each other, it takes $\xi_{\perp}^2$ steps to move over the distance of the
perpendicular correlation length. This mechanism destroys correlations of the
order $\xi_\perp$ in the spin orientation after $ \xi_{\perp}^2\exp\left(2\beta
J_\parallel\right)$ time steps.
We conclude
\begin{equation}
 \xi_{\parallel} \propto
\exp\left(8\beta J_\perp + 2 \beta J_\parallel\right)\propto
\xi_{\perp}^{2+J_{\parallel}/2J_{\perp}}\,,
\label{z}
\end{equation}
leading to $z=2+J_{\parallel}/2J_{\perp}$.
Depending on the ratio between the
two coupling constants, the critical exponent $z$ can assume any value larger
than two. Since $z$ is larger than one, the parallel correlation length becomes
much larger than the perpendicular correlation length when the temperature is
low. On the other hand, we can expect that for high temperatures the direction
that has the stronger coupling has the larger  correlation length. Thus, if
$J_{\parallel} < J_{\perp}$, the two correlations lengths must cross
at some temperature.

The above intuitive argument (see also \cite{cor81} for a similiar argument
in the context of one-dimensional Ising models)
 does not consider the creation and annihilation of
pairs of domain walls, which play an important role in the dynamics of the
system. In the following, we therefore derive the exponent $z$ from the
complete dynamics of the domain walls. We will see that models with different
$J_\parallel$ but the same $J_\perp $ are equivalent, apart from an overall
time scale proportional to $\exp(-2\beta J_\parallel)$. We will also see that
 these models are closely related to the Glauber model \cite{glau} describing
the dynamics of a one-dimensional Ising spin chain that is known to have a
dynamical critical exponent $z=2$. In the Glauber model, it does not cost any
energy to move a domain wall, and therefore this motion does not become slower
with decreasing temperature.

\begin{figure}
\narrowtext
\centerline{\epsfysize=3in 
\epsffile{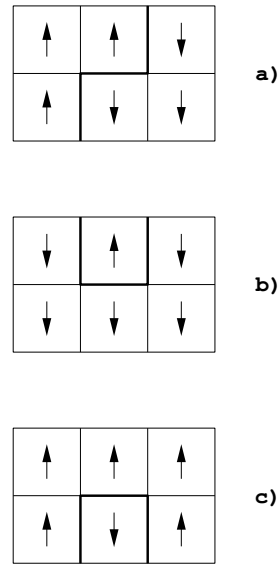}}
\vskip 0.5cm
\caption{The three elementary processes in domain wall dynamics: a) motion of a
wall b) creation of a pair of walls, c) annihilation of a pair of walls.}
\label{walls}
\end{figure}

The domain wall dynamics consist of three different elementary processes:
motion to the right or left by one site, creation of a pair of neighboring
domain walls, and annihilation of a pair of neighboring domain walls (see
Fig.~\ref{walls}). 
These three processes occur with probabilities
$\exp(-2\beta J_\parallel)$, $\exp[-2\beta(2 J_\perp + J_\parallel)]$, and
$\exp(-2\beta J_\parallel) /[\exp(-2\beta J_\parallel)+ \exp(-4\beta J_\perp)]$
respectively. These probabilities are obtained by dividing the Boltzmann
factor for the considered event by the sum of the Boltzmann factors of all
possible events at that place, and retaining only the leading contributions.
The neglected terms become vanishingly small compared to the leading terms in
the limit of zero temperature. In the denominator of the third expression,
the Boltzmann factor for domain wall annihilation can be larger or smaller
than the Boltzmann factor for the two domains walls staying at their place,
depending on the values of the coupling constants. We therefore keep both
contributions.
For $J_\parallel > 2 J_\perp$, the three
probabilities are all very small and can therefore be interpreted as rates. If
we extract a time scale $\exp(-2\beta J_\parallel)$, these rates become $1$,
$\exp(-4\beta J_\perp )$, and $\exp(4\beta J_\perp )$, independently of the
value of $J_\parallel$, showing that models with different $J_\parallel$ differ
only by the overall time scale. For  $J_\parallel < 2 J_\perp$, the situation
is slightly different: A pair of neighboring domain walls is annihilated almost
with probability one during a single time step, while this process occurs for
$J_\parallel > 2 J_\perp$ on an average only after $\exp[2\beta(J_\parallel - 2
J_\perp)]$ time steps. However, in both cases this process is very fast
compared to the time it takes to move a domain wall or to create a pair of
domain walls. Additionally, the probability that a pair of neighboring domain
walls ultimately escapes annihilation by increasing their distance to two is
$2\exp(-4\beta J_\perp)$ in both cases.
This shows that (apart from the overall
time scale) models with $J_\parallel < 2 J_\perp$  are equivalent to those with
$J_\parallel > 2 J_\perp$.
(A short calculation shows that they are also equivalent to models with
$J_\parallel = 2 J_\perp$.)
This equivalence becomes even more obvious when we
describe the model in terms of creation and annihilation of pairs of domain
walls at distance two, and of domain wall motion. Domain walls at distance two
are created by first flipping one spin (creating a pair at distance one) and
then flipping one of its neighbors, which happens in all cases with a rate
$2\exp(-2\beta J_\parallel - 8\beta J_\perp)$. (Strictly speaking, the pair at
distance one can not only be generated by flipping a spin in a homogeneous
domain, but also by the encounter of two domain walls that have been further
apart before. However, it can easily be calculated that the number of pairs at
distance one generated by this process is smaller by a factor of $\exp(-4\beta
J_\perp)$ than the number of those generated by flipping a spin in a
homogeneous domain.)
Domain walls at distance two decrease their distance to one with a rate
$2\exp(-2\beta J_\parallel)$, from where they annihilate almost certainly. If
we extract a factor $\exp(-2\beta J_\parallel)$ from the time scale, motion of
domain walls, creation of pairs of domain walls at distance two, and
annihilation of pairs of domain walls at distance two, occur with the rates
$1$, $2\exp(-8\beta J_\perp) = \xi_\perp^{-2}/2$, and $2$, irrespective of the
value of $J_\parallel$. Since we can neglect the interference of pairs of
domain walls at distance one with any of these three processes, the dynamics of
the model is completely described by these three rates.

To complete this discussion, we show that the dynamics of our model are
essentially equivalent to the dynamics of the Glauber model. In the Glauber
model, the rate with which a spin at site $i$ flips is given by \cite{glau}
\begin{equation}
W(\sigma_i) =
{\alpha\over2}\left[1-{\gamma\over2}\sigma_i\left(\sigma_{i+1}+\sigma_{i-1}
\right)\right]\,,\label{glau}
\end{equation}
with $\gamma = \tanh(2\beta J) \simeq 1 - 2\exp(-4\beta J) \simeq 1 -
\xi^{-2}/2$ at low temperatures. The parameter $\alpha$ sets the time scale and
is independent of temperature. Motion of a domain wall, creation of a pair of
neighboring domains walls, and annihilation of a pair of neighboring domain
walls occur with the rates $\alpha/2$, $\alpha\xi^{-2}/4$, and $\alpha$
respectively. If we set $\alpha = 2\exp(-2\beta J_\parallel)$, we find the same
rates as for our model, with the only difference that creation and annihilation
of pairs of domain walls happens at distance one instead of distance two. This
difference is obviously irrelevant at low temperatures, where domain walls must
diffuse over large distances of the order of the correlation length before they
encounter each other. Reducing this distance by one has no noticeable effect on
the dynamics of the system. Combining the critical exponent $z=2$ of the
Glauber model with the temperature dependence of the time scale in our model,
we find Eq.~(\ref{z}).

\begin{figure}
\narrowtext
\centerline{\epsfysize=1in 
\epsffile{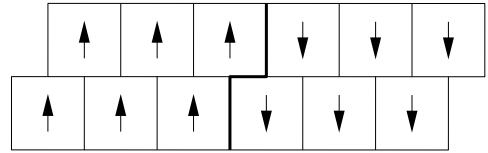}}
\caption{A different lattice structure that allows domain wall motion at no
additional
energy.}
\label{neu}
\end{figure}

Nonuniversal dynamical critical exponents are also known from other 
one-dimensional Ising systems. An Ising spin chain with two 
different coupling constants $J_1$ and $J_2$, with  $J_1 < J_2$, has an 
exponent $z = 1+ J_2/J_1$ \cite{dro86,dro87}. 
Finally, we note that the result Eq.~(\ref{z}) depends strongly on the lattice
structure of the
system. When the cells of the lattice do not lay on top of each other but
are shifted by half a lattice constant as in Fig.~\ref{neu}, the motion of a
domain wall does not cost any energy, and we retrieve the exponent $z=2$.

\section{General Dimensions}

The overall probability for encountering a particular sequence of layers is
given by
\begin{equation}\label{Pall}
P\left( \alpha_1, \alpha_2,\cdots, \alpha_N\right)=
W_{ \alpha_1 \alpha_2}W_{ \alpha_2 \alpha_3}\cdots W_{ \alpha_{N-1} \alpha_N}.
\end{equation}
Substituting from Eq.(\ref{W}), it follows that the numerator of the above
expression is simply the the Boltzmann weight for a regular ferromagnetic
nearest neighbor Ising model. The denominator is the product of factors
\begin{eqnarray}
Z_\alpha&\equiv &\exp\left(+\beta \tilde E_\alpha\right)=\sum_\gamma
\exp\left[-\beta \left(V_{\alpha\gamma}+E_{\gamma}\right)\right] \label{Za}\\
&=& \sum_\gamma \exp\left[\beta J_\parallel\sum_i \sigma_{i}^{(\alpha)}
\sigma_i^{(\gamma)}
+\beta J_\perp \sum_{<ij>}\sigma_i^{(\gamma)}\sigma_j^{(\gamma)}
\right],\nonumber
\end{eqnarray}
one for each layer. After the summation is performed, $\tilde E_\alpha$
introduces
additional interactions within layer $\alpha$. These interactions are in
general
non-local, and involve multiple spins, rendering the problem highly
anisotropic,
and non-trivial.

We first estimate $\tilde E_\alpha$ by a high temperature expansion.
For each Ising bond appearing in Eq.(\ref{Za}), we can write
\begin{equation}
\exp\left(\beta J\sigma_i\sigma_j\right) = c\left(1+ t \sigma_i\sigma_j\right),
\label{highT}
\end{equation}
where $c=\cosh(\beta J)$, and $t =\tanh(\beta J)$ is the small parameter for
the
high temperature expansion. After summing over the spins in layer $\gamma$,
the surviving contributions to $Z_\alpha$ are represented graphically by paths
connecting pairs of spins in layer $\alpha$. For hypercubic layers of
coordination
number $2d$,
\begin{eqnarray*}
Z_\alpha &=& \left( c_\parallel c_\perp^d \right)^N
\sum_{\{\sigma^{(\gamma)}\}}\\
&&\prod_i \left(1+t_\parallel\sigma_i^{(\alpha)}\sigma_{i}^{(\gamma)}\right)
\prod_{<ij>}\left(1+t_\perp\sigma_i^{(\gamma)}\sigma_{j}^{(\gamma)}\right)\\
&=& \left( c_\parallel c_\perp^d \right)^N
\left[ 1 + \sum_{<ij>}t_\parallel^2 t_\perp
\sigma_i^{(\alpha)}\sigma_j^{(\alpha)}
+\cdots\right]\\
&\approx&\left( c_\parallel c_\perp^d \right)^N
\exp\left( +t_\parallel^2 t_\perp\sum_{<ij>}
\sigma_i^{(\alpha)}\sigma_j^{(\alpha)}
+\cdots \right).
\end{eqnarray*}
After inversion, the leading effect of the denominators is to {\it reduce} the
in--plane couplings to $\beta J_\perp-t_\parallel^2 t_\perp$.
At higher orders,  anti-ferromagnetic bonds of strength $t_\parallel^2
t_\perp^s$,
for spins at a distance $s$, and multiple spin interactions are also generated.

A similar high temperature expansion results in the
parallel and perpendicular correlation functions
\begin{displaymath}
g_\perp(l)\simeq t_\perp^l\left\{1+ {l(l+1)\over 2}\left[2(d-1)t_\perp^2 +
t_\parallel^2\right] + \cdots\right\},
\end{displaymath}
and
\begin{displaymath}
 g_\parallel(l)\simeq t_\parallel^l [1+ d l(l+1) t_\perp^2  + \cdots]\,.
\end{displaymath}
The weakening of the in--plane bonds results in a reduced correlation length
perpendicular to the growth direction. In fact, such a weakening is precisely
what is needed to make the correlations within each layer appear as
$d$--dimensional, despite the $(d+1)$--dimensional connectivity of the overall
system of spins. The anisotropy in correlations is amplified at lower
temperatures,
and ultimately they diverge with different exponents at the critical point.

At very low temperatures, there are typically few unaligned neighbors,
and in a low temperature expansion, Eq.(\ref{Za}) is evaluated as
\begin{equation}\label{Zalt}
Z_\alpha\approx  \exp\left( N\beta J_\parallel+\beta J_\perp
\sum_{<ij>}\sigma_i^{(\alpha)}\sigma_j^{(\alpha)} \right).
\end{equation}
(The spins in layer $\gamma$ are assumed to be completely aligned to
those in layer $\alpha$ in the configuration of lowest energy.)
Thus at this order, the effective in--plane couplings are $\beta J_\perp-
\beta J_\perp\approx 0$! The physical meaning of this result is that at
low temperatures each new layer essentially repeats the configuration
of the previous layer. The boundaries between different domains have
very small mobility (of order of $\exp\left( -\beta J_\parallel \right)$).
It is this reduced mobility that leads to the unusual critical properties of
the $1+1$ dimensional model discussed earlier.
In higher dimensions, the phase transition occurs at finite temperature, and
the
mobility of  a domain wall should be finite in the neighborhood of the critical
point. Since our model satisfies detailed balance, and there is no local
conservation of spins from one layer to the next, we expect it to be in the
universality class of an Ising system with model A dynamics. The
correlations parallel the growth direction diverge therefore with an exponent
$z\nu$,
where $z$ is  the (Glauber) dynamic exponent of the $d$-dimensional Ising
model.

\section{Conclusions}

In summary, we have studied a model for thin film growth which explicitly
includes the relaxation and diffusion of the atoms on the top surface.
Since there is no further relaxation inside the film, the bulk configuration
reflects the time history of fluctuations at the surface.
The resulting correlations are highly anisotropic: within each layer,
they correspond to the equilibrium on a $d$--dimensional system,
while the behavior parallel to the growth direction reflects the dynamics of
such a system.  These results are similar to previous (cellular automata
inspired) models which include no surface relaxation. This suggests
that a wider universality is present: as long as relaxation stops after a
finite
number of layers from the top surface, the above conclusions should hold.
One exception is for $1+1$ dimensional systems, which due to the
vanishing mobility of domains at zero temperature, exhibit non--universal
behavior.

A crucial assumption of the model is that the surface layer has sufficient time
to equilibrate, before it is completed and incorporated in the bulk.
This requires that any surface relaxation times $\tau_R$ should be
less than the time for the growth of a layer $\tau_G$. The latter is
controlled by the speed of deposition. The former time, however, depends
on the actual dynamics at the surface. It is important to emphasize that
$\tau_R$ has nothing to do with the Glauber dynamics that determine
the correlations in growth direction. If the main relaxation mechanism
is diffusion of particles on the surface,  $\tau_R$ will be quite long, and
diverge close to the critical point as $\tau_R\propto \xi_\perp^{z_c}$,
where $z_c>z$ is the exponent for conservative (model B) dynamics.
On the other hand, $\tau_R$ can be considerably reduced if the surface
particles are allowed to desorb into the gas. By fast mixing of the vapor,
it is then possible to achieve a $\tau_R$ that is independent of the
dynamics at the surface.

If the inequality $\tau_R<\tau_G$ is violated, the most likely scenario is
that in--plane relaxations are frozen at some scale, resulting in
configurations similar to those obtained in a rapid quench.
The situation is further complicated in the low temperature
phase. Phase separation during deposition results in the coarsening of domains,
as observed in experiments on Al-Ge \cite{ada93}. Computer simulations of this
process generate at low growth speed domains that extend into filaments along
the growth
direction \cite{ada93}, a result also obtained from analytical calculations
\cite{atz92}. A similar study for ``antiferromagnetic'' ordering during film
growth has not yet been performed to our knowledge. Since the dynamics for
antiferromagnetic ordering do not require long-range transport of particles,
large correlated domains can be formed during finite time, given large
correlations in the layer below. We therefore expect that a small finite growth
velocity does not destroy the phase transition of ordering systems, but only
shifts the critical temperature to a lower value.

All considerations so far are based on the assumption that the surface of the
film is flat. It is likely that surface roughness and formation of domains will
affect each other.
It is thus quite interesting to explore the interconnections between
roughness and phase transitions in composite film growth.
Another complication that exists in recent experiments on growth of
Co-Pt alloys \cite{roo95}, is that one of the two components
is magnetic. The additional magnetic interactions provide yet another
twist to this interesting problem.

We thank Y. Bar-Yam, F. Hellman, and A. Stella for helpful discussions.
BD is supported  by the  Deutsche
Forschungsgemeinschaft (DFG) under Contract No.~Dr 300/1-1.
MK acknowledges support from NSF grant number DMR-93-03667.

\end{multicols}
\end{document}